\documentstyle[prl,preprint,aps,graphicx,amsfonts,amssymb,
amsmath]{revtex}
\begin{document}
\draft
\preprint{}
\title{Intrinsic Resistivity via Quantum Nucleation of Phase 
Slips in a One-Dimensional Josephson Junction Array}
\author{T. Inoue, M. Nishida and S. Kurihara}
\address{
Department of Physics, Waseda University, 3-Okubo, Shinjuku-ku,
Tokyo 169-8555, Japan}
\date{\today}
\maketitle
\begin{abstract}
 The resistivity of a one-dimensional Josephson junction array at zero 
 temperature is calculated by estimating the nucleation rate of 
 quantum phase slips. We choose a certain collective coordinate 
 which describes the nucleation process and estimate the corresponding 
 effective mass.
 In the strong coupling regime, where Josephson coupling energy 
 exceeds the charging energy, we calculate the nucleation rate by means of 
 WKB method. 
 Our estimation is in good  agreement with recent experimental data. 
 The superconductor-insulator transition point is also discussed.
\end{abstract}

 Quantum fluctuations have much influence on 
 the properties of one-dimensional systems at low temperatures. For example, 
 in extremely thin superconducting wires which can be regarded as nearly 
 one-dimensional, the quantum fluctuations of the phase of the order parameter 
 nucleate phase slips, which result in finite resistance at zero 
 temperature \cite{giordano94}. 
 Similar effects are expected for a one-dimensional Josephson junction (1D JJ) 
 array. In a 1D JJ array, the phase difference 
 and charge difference across a junction are canonically 
 conjugate variables, so the charging effect is the origin of the quantum 
 fluctuation of phase. 
 When the system size becomes small and the charging energy large enough, 
 nucleation occur frequently and at some critical point, the quantum phase 
 transition from a phase-coherent (superconducting) state to a charge-ordered 
 (insulating) state occurs. 
 Theoretically, this Superconductor-Insulator (S-I) transition in a 
 1D JJ array can be understood as some version of Kosterlitz-Thouless 
 transition in a (1+1) dimensional classical spin model, extra 
 dimension being imaginary time \cite{bradley84,sondhi96}. 

 Recently, such S-I transition was experimentally investigated in 
 detail with 1D JJ arrays having different numbers of superconducting grains 
 \cite{haviland98}. In that work, the S-I transition was actually observed for 
 each $N$ near the point predicted by Kosterlitz-Thouless theory, and what is 
 also remarkable, it was found that arrays have finite resistance at zero 
 temperature even in the superconducting side. 
 It is probable that these intrinsic resistance in the superconducting side 
 are caused by the quantum nucleation of phase slips, as in the case 
 of thin superconducting wires. 
 In order to confirm this interpretation, we study the quantum nucleation 
 process of a phase slip and calculate the nucleation rate in this Letter. 
 The S-I transition point is also discussed.

 To begin with, we consider a linear array of $N (\gg1)$ superconducting grains  embedded in an insulator at zero temperature. 
 The grains are assumed to be small compared with bulk coherence length, so 
 that the state of $i$th grain can be described by a single phase 
 $\theta_i$ and number $n_i$ of Cooper pairs on the $i$th grain. 
 The Euclidean action and Hamiltonian are 
 \begin{equation}
  S_{\rm E}[ \left\{n_i,\theta_i\right\} ] =
  \int d\tau\,
  \left\{\sum_i^N i\hbar n_i \frac{\partial\theta_i}{\partial \tau}+H
  [\left\{n_i,\theta_i\right\}] \right\},
  \label{eq:action}
 \end{equation}
 \begin{multline}
	H[ \left\{n_i,\theta_i\right\} ] = \sum_{i=1}^{N-1}\frac{1}{2}E_{\rm C}
	\left(\frac{n_{i+1}-n_i}{2}\right)^2 \\
	+ \sum_{i=1}^N\frac{1}{2}E_{\rm G}\left(n_i-\bar{n}\right)^2 
	- \sum_{i=1}^{N-1}E_{\rm J}\cos(\theta_{i+1}-\theta_i). 
  \label{eq:Hamiltonian}
 \end{multline}
 In Eq. (\ref{eq:Hamiltonian}), we take into account two kinds of charging 
 energies, $E_{\rm G}$ and $E_{\rm C}$, the former is between a grain and the 
 ground and the latter between nearest neighbor grains. 
 $E_{\rm J}$ is Josephson coupling energy and $\bar{n}$ is the average number 
 of Cooper pairs on a grain.  Although the normal component is not explicitly 
 mentioned here, the property of the system largely depends on the value of 
 normal tunnel resistivity $R_{\rm T}$ \cite{chakravarty82,moore82,leggett87}. 
 In this Letter, we restrict our attention to the superconducting regime, 
 so $R_{\rm T}$ is assumed to be smaller than the quantum resistance 
 $R_{\rm Q}=\frac{h}{4e^2}\sim 6.45 {\rm k\Omega}$.

 A supercurrent state of the above Hamiltonian is metastable, even at zero 
 temperature. Such a state corresponds to a local minimum of the energy 
 functional of $n_i$ and $\theta_i$, and definitely decay to lower local 
 minima, due to quantum tunneling. When the effective 
 inertial mass of the phase slip is not greatly dependent on the tunneling 
 path, the most probable tunneling path will be through the valley, passing 
 through two local minima and one saddle point in the functional space.
 In this situation, the wave function which describes the phase slip 
 process is determined so as to minimize the energy functional at 
 each stage of the tunneling process. 
 Of many tunneling paths, we adopt only paths which describe slips of $2\pi$. 
 Other paths of $4\pi,6\pi,\ldots$ can be omitted because of their much higher 
 tunnel barriers. 

 To determine the shape of the path, we temporarily fix the phase slip center 
 between the $i$th and $(i+1)$th grain and restrict the average value of 
 supercurrent, 
 so add the following two terms in Hamiltonian Eq. (\ref{eq:Hamiltonian})
 \begin{equation}
  \lambda_1 \left( \theta_{j+1} - \theta_j - q \right) 
  + \lambda_2 \left\{ \sum_{i=1}^{N-1}
  \left( \theta_{i+1} - \theta_i - \frac{2\pi n}{N-1}\right) \right\}, 
  \nonumber
  \label{eq:boundary condition}
 \end{equation}
 where $q$ is the "phase jump" at the phase slip center satisfying 
 $-\pi<q<\pi$. The "winding number of the phase" $n$ is a real 
 number, but not necessarily an integer. 
 The extremal solution of the energy functional with additional terms is  
	\begin{eqnarray}
	 n_i &\equiv& \bar{n}, \hspace{5mm}
	 \theta_{i+1} - \theta_i = \kappa = \frac{2\pi n-q}{N-2} \hspace{5mm} 
	 \left( i\neq j \right), \nonumber \\
	 n_j &\equiv& \bar{n},\hspace{5mm} \theta_{j+1} - \theta_j = q .
	 \label{eq:E-minimum configuration}
	\end{eqnarray}
 When we take the phase jump $q$ 
 as the collective coordinate describing each 
 stage of the tunneling process, Eq. (\ref{eq:E-minimum configuration}) is 
 the energy minimum configuration with given $q$ and $n$. 
 When $q=\kappa$, Eq. (\ref{eq:E-minimum configuration}) gives the 
 local minimum solution $\left\{n_{{\rm m},i},\theta_{{\rm m},i}\right\}$:
 \begin{equation}
  n_{{\rm m},i} \equiv \bar{n}, \hspace{0.5cm}
  \theta_{{\rm m},i+1} - \theta_{{\rm m},i} \equiv \kappa_n = 
  \frac{2\pi n}{N-1}, 
  \label{eq:local minimum solution}
 \end{equation}
 and when $q=\pi-\kappa$, the saddle point solution 
 $\left\{n_{{\rm s},i},\theta_{{\rm s},i}\right\}$: 
 \begin{eqnarray}
  n_{{\rm s},i} &\equiv& \bar{n}, \hspace{0.5cm}
  \theta_{{\rm s},i+1} - \theta_{{\rm s},i} = \kappa = 
  \frac{\left( 2n-1 \right)\pi}{N-3} 
  \hspace{5mm} \left( i \neq j \right), \nonumber \\
  n_{{\rm s},j} &\equiv& \bar{n},\hspace{5mm} \theta_{{\rm s},j+1} - 
  \theta_{{\rm s},j} = \pi-\kappa.
  \label{eq:saddle point solution}
 \end{eqnarray}
 Now the wave function is constructed which interpolate two local minimum 
 points in the functional space. 

 Hereafter we adopt $\kappa$, rather than $q$, as the collective coordinate, 
 because using $\kappa$ allows us to consider 
 the whole of the tunneling process while $q$ doesn't.
 The potential energy is now a function of the collective coordinate $\kappa$,
 \begin{equation}
  V \left( \kappa \right) = -E_{\rm J}\left[ \left( N-2 \right)\cos \kappa 
 + \cos \left\{ 2\pi n -\kappa \left( N-2 \right) \right\} \right].
  \label{eq:potential}
 \end{equation}

\begin{figure}[htpb]
\begin{center}
\includegraphics[keepaspectratio=true, width=\textwidth]%
{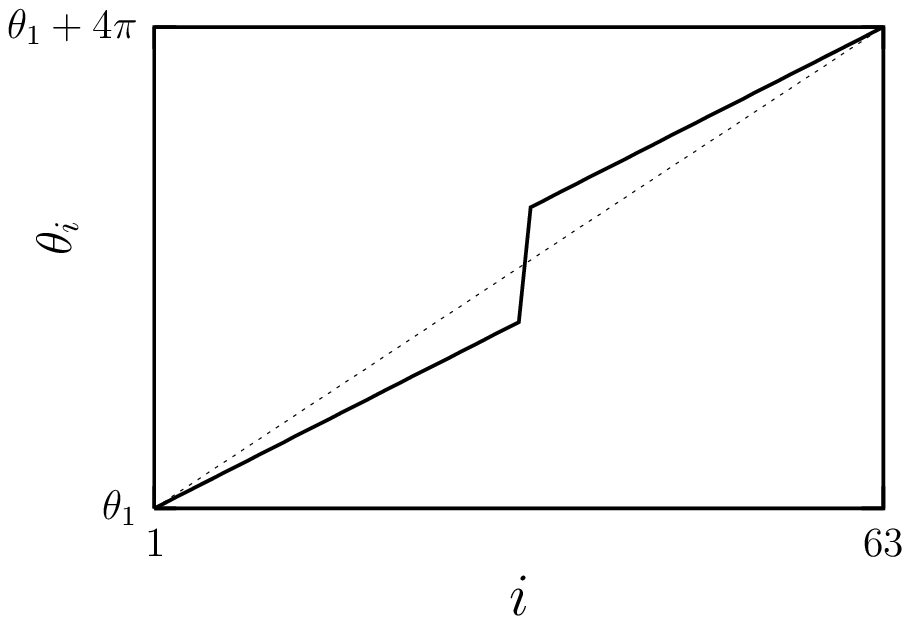}
\caption{The phase of the interpolation function 
Eq. (\ref{eq:E-minimum configuration}) for the case $N=63, n=2$ and $q=3$ 
is plotted (solid line). The dotted line depicts the local minimum solution Eq. (\ref{eq:local minimum solution}).}
\label{fig:interpolation}
\end{center}
\end{figure}

 Next, we discuss the zero-point fluctuation of the collective
 coordinate $\kappa$ around the local minimum. 
 As already mentioned, $n_i$ and $\theta_i$ are canonically conjugate 
 variables, therefore the field necessarily fluctuates around the 
 local minimum configuration Eq. (\ref{eq:local minimum solution}) and 
 a certain mode of this fluctuations corresponds to the zero-point fluctuation 
 of $\kappa$ which nucleates phase slips. 
 In order to estimate frequencies of collective modes, we add the 
 fluctuation terms $\left\{\delta n_i,\delta \theta_i\right\}$ to the local 
 minimum configuration Eq. (\ref{eq:local minimum solution}),  
 substitute it into the Euclidean action, Eq. (\ref{eq:action}), and retain the  terms up to the second order of $\delta n_i$ and $\delta \theta_i$, 
 then we obtain 
	\begin{equation}
	 \delta S_{\rm E} 
	 = \sum_{i=1}^N \int {\rm d\tau}
	 \left(\begin{array}{cc}
	  \delta n_i & \delta \theta_i 
	  \end{array}\right)
	 \mbox{\boldmath $G$}^{-1}
	 \left(\begin{array}{c}
	  \delta n_i \\
	  \delta \theta_i 
	 \end{array}\right).
	 \label{eq:fluctuation to second order} 
	\end{equation}
 Here, $\boldsymbol{G}$ is the Green's function for $\delta n_i$ and 
 $\delta \theta_i$, and Fourier component of $\boldsymbol{G}^{-1}$ is given by 	\begin{equation}
	 \boldsymbol{G}^{-1} = 
	 \left(\begin{array}{cc}
	 \frac{1}{2}E_{\rm G}+\frac{1}{2}E_{\rm C}\sin^2\frac{ka}{2}
	 & \frac{\hbar}{2}\omega_n \\
	 -\frac{\hbar}{2}\omega_n &
	 2E_{\rm J}\cos\left(\frac{2\pi n}{N-1}\right)\sin^2\frac{ka}{2}
	 \end{array}\right),
	\end{equation}
 where $a$ is the distance between nearest neighbor grains. 
 The poles of $\boldsymbol{G}$ give collective modes, whose dispersion 
 relation is 
	\begin{equation}
	 \hbar\omega_k = 2\sqrt{
	 \left\{E_{\rm G}+E_{\rm C}\sin^2\frac{ka}{2}  \right\}
	 E_{\rm J}\cos\left(\frac{2\pi n}{N-1}\right)
	 \sin^2\frac{ka}{2} }.
	 \label{eq:dispersion}
	\end{equation}
 Fig. \ref{fig:interpolation} suggests that a phase slip has a local character,  thus as the effective frequency of the zero-point oscillation of $\kappa$, 
 we can extract from Eq. (\ref{eq:dispersion}) the mode whose wave length 
 is $a$: 
	\begin{equation}
	 \hbar\omega_{{\rm eff}}(\kappa_n) = 
	 \hbar\omega_{k=\frac{\pi}{a}} = 2\sqrt{
	 \left(E_{\rm G}+E_{\rm C}\right)
	 E_{\rm J}\cos\left(\frac{2\pi n}{N-1}\right)}.
	 \label{eq:zero point frequency}
	\end{equation}
 We have already determined the potential 
 energy, Eq. (\ref{eq:potential}), so the effective inertial mass of 
 $\kappa$ can be estimated from the frequency of the collective mode and 
 the curvature of the potential energy at the local minimum.
 In this procedure, we assume that the inertial mass is not greatly 
 dependent on the collective coordinate, and approximate it to the value 
 at local minimum.
 From Eq. (\ref{eq:potential}) the potential curvature at the local minimum 
 $\kappa \sim \kappa_n$ is 
	\begin{equation}
	 V''(\kappa_n)=E_{\rm J}\left( N-2 \right)\left( N-1 \right) 
	 \cos\left(\frac{2\pi n}{N-1}\right),
	 \label{eq:curvature}
	\end{equation}
 where the prime denotes differentiation with respect to $\kappa$.
 We can derive the effective inertial mass for $\kappa$ 
 from Eq.\ (\ref{eq:zero point frequency}) and Eq.\ (\ref{eq:curvature}), 
	\begin{equation}
	 M_{\kappa}(n) = \frac{V''(\kappa_n)}{\omega_{\text{eff}}^2(n)}
	 = \frac{\hbar^2}{4}
	 \frac{\left(N-2 \right) \left(N-1 \right)}{E_{\rm G}+E_{\rm C}}.
	 \label{eq:effective mass}
	\end{equation}

 We now consider the tunneling problem from the local minimum $\kappa_n$ 
 to the next lower local minimum $\kappa_{n-1}$.
 According to WKB formula, the tunneling rate 
 $\Gamma(\kappa_n\rightarrow\kappa_{n-1})$ is given by 
	\begin{multline}
	 \Gamma(\kappa_n\rightarrow\kappa_{n-1}) 
	 = \frac{\omega_{\text{eff}}(\kappa_n)}{2\pi}
	 \exp\left\{ -\frac{2}{\hbar}S(\kappa_n) \right\}, \\
	 S(\kappa_n) = \int_{\kappa_n^-}^{\kappa_n^+} 
	 \sqrt{2 M(\kappa_n)
	 \left( V(\kappa)- \frac{1}{2}\hbar\omega_{k=\frac{\pi}{a}}\right)},
	 \label{eq:tunneling rate}
	\end{multline}
 where $\kappa_n^-$ and $\kappa_n^+$ are the classical turning points, 
 located between $\kappa_{n-1}$ and $\kappa_n$. 
 Here $\kappa_n^+$ doesn't coincide with $\kappa_n$ because we adopt not 
 the instanton method, but WKB method which explicitly includes the zero point 
 energy in the action. This point will be discussed later.
 Besides, we implicitly assume the existence of the environments, so that the 
 energy is conserved before and after the tunneling. 

 In the limit of large $N$ and small bias current, we can reduce the 
 action in Eq. (\ref{eq:tunneling rate}) up to the first order of a small 
 parameter $\frac{n}{N}$, to obtain  
	\begin{multline}
	 S(\kappa_n) \sim 2\hbar J \times \\
	 \left\{ \begin{array}{r}
	 2\left( E\left(\frac{\pi}{2},\sqrt{1-\frac{1}{2J}} \right)
	 -\frac{1}{2}F\left(\frac{\pi}{2},\sqrt{1-\frac{1}{2J}} \right) J 
	 \right) \\ 
	 - \pi^2 \frac{n}{N} 
	 F\left(\frac{\pi}{2},\sqrt{1-\frac{1}{2J}} \right) \end{array} \right\},
	 \label{eq:n/N extension of action}
	\end{multline}
 where $J=\sqrt{\frac{E_{\rm J}}{E_{\rm G}+E_{\rm C}}}$ is the coupling 
 parameter representing the strength of the phase coherence. $F$ and $E$ 
 denotes the first and second kind of elliptic integrals respectively. 

 From the tunneling rate, we can derive an expression for the resistivity. 
 Combining Josephson's relation 
	\begin{equation}
	 \frac{d\theta_{{\rm tot}}}{dt} = \frac{2e}{\hbar}V
	 \label{eq:Josephson's relation}
	\end{equation}
 and the equation of motion of the phase
	\begin{equation}
	 \frac{d\theta_{{\rm tot}}}{dt} = 2\pi 
	 \Gamma(\kappa_n\rightarrow\kappa_{n-1}), 
	 \label{eq:equation of motion of the phase}
	\end{equation}
 we conclude that the voltage at metastable state $\kappa\sim\kappa_n$ is 
 \begin{equation}
 V(\kappa_n) = \frac{\hbar \pi}{e}\Gamma(\kappa_n\rightarrow\kappa_{n-1}),
 \label{eq:voltage}
 \end{equation}
 where $\theta_{{\rm tot}}$ denotes the total difference of the phase 
 throughout the array.
 The zero bias resistivity $R$ is defined as follows:
	\begin{equation}
	 R = \frac{\partial V(\kappa_n)}
	 {\partial I(\kappa_n)} \bigg|_{\frac{n}{N}\rightarrow 0},
	 \label{eq:zero bias resistivity}
	\end{equation}
 where $I(\kappa_n)$ is the applied current equal to $E_{\rm J}\sin \kappa_n$. 

 In the limit $N \rightarrow \infty$, we can actually calculate the zero 
 bias resistivity in the analytical form. From 
 Eq. (\ref{eq:tunneling rate}), (\ref{eq:n/N extension of action}), 
 (\ref{eq:voltage}) and (\ref{eq:zero bias resistivity}), $R$ is obtained in 
 the following form:
	\begin{multline}
	 R = 2R_{\rm Q} \times 
	 F\left(\frac{\pi}{2},\sqrt{1-\frac{1}{2J}} \right) \times \\
	 \exp \left[ 
	 -8E\left(\frac{\pi}{2},\sqrt{1-\frac{1}{2J}} \right) J 
	 +4F\left( \frac{\pi}{2},\sqrt{1-\frac{1}{2J}} \right) \right]. 
	 \label{eq:resistivity}
	\end{multline}
 Note that $J$ contains both $E_{\rm G}$ and $E_{\rm C}$. 
 Many theoretical works have treated only the limiting case in which either 
 $E_{\rm G}$ or $E_{\rm C}$ is absent, while our derivation 
 includes the case in which the two charging energies are comparable. 
 Incidentally, it is reported that the resistivity is nonlinearly dependent on 
 the applied current in the case of superconducting wires \cite{saito89}, but 
 that might be caused by applying, instead of WKB method, the instanton method 
 \cite{coleman77} which doesn't explicitly take into account the zero point 
 energy in the action. This is the reason why we adopt WKB method, instead of 
 the instanton method.

\begin{figure}[htpb] 
\begin{center}
\includegraphics[keepaspectratio=true, width=\textwidth]%
{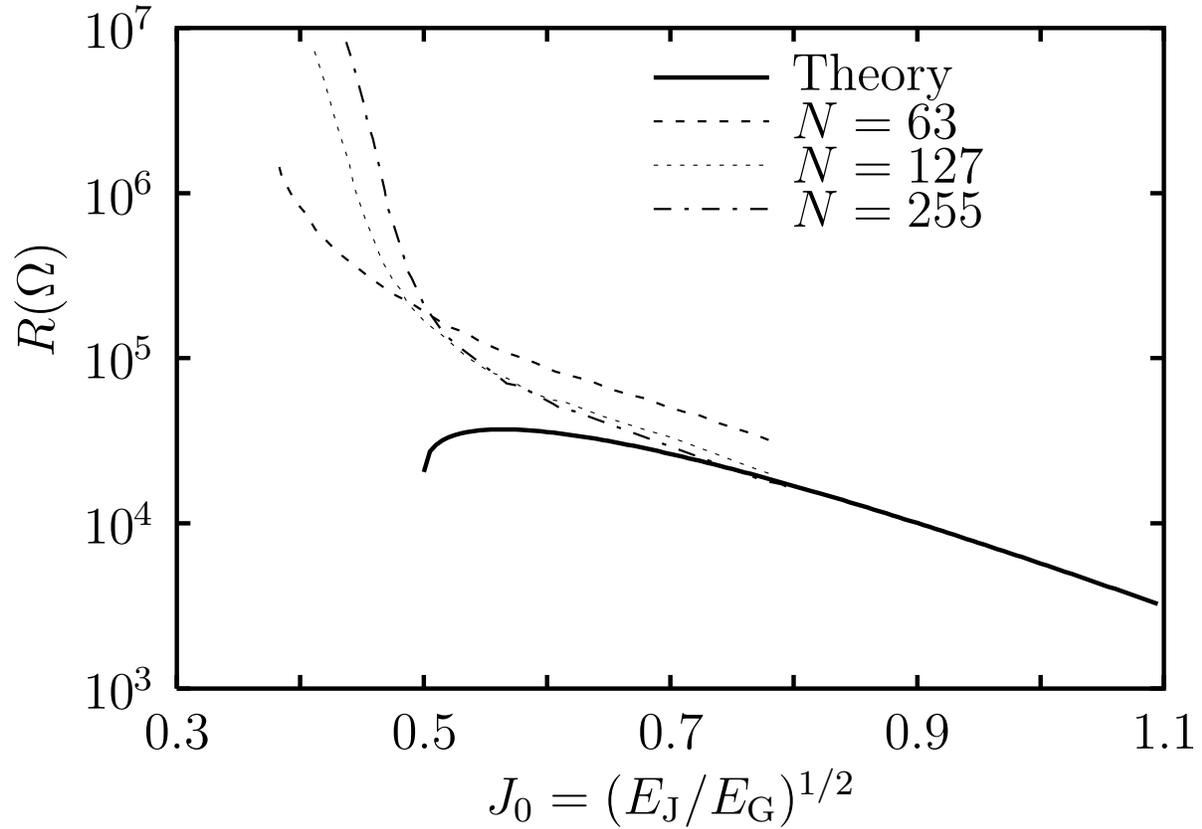} 
\caption{The zero bias resistivity plotted against the coupling parameter 
 $J_0$. In the superconducting side $(J_0\gtrsim0.5)$, our theoretical value 
 is in good agreement with the experimental data, especially with the 
 ones for $N=127$ and $N=255$.}
\label{fig:resist}
\end{center}
\end{figure}

 The resistivity Eq. (\ref{eq:resistivity}) is shown in Fig. \ref{fig:resist} 
 compared with the experimental result. 
 Because the experiment is in the limiting case where 
 $E_{\rm G}\gg E_{\rm C}$, the data are plotted against the reduced parameter 
 $J_0=\sqrt{\frac{E_{\rm J}}{E_{\rm G}}}$. 
 Without any fitting parameter, our theory and the experiment show good 
 agreement asymptotically in the superconducting side $(J_0\gtrsim0.5)$. 
 However WKB approximation, on which our theory is based, is expected to be 
 valid in the strong coupling limit $(J_0\gg1)$, so more experimental data in 
 the stronger coupling regime $(J_0>0.8)$ are necessary to confirm our theory. 
 Further, the data of arrays in which $E_{\rm G}$ and $E_{\rm C}$ 
 are comparable are also hoped for. 

 We can also estimate the S-I transition points for various number of 
 grains $N$. In a strict sense, there is no critical point for a finite $N$ 
 because there is no sharp phase transition in a finite size system. But 
 instead we can consider some critical region within which the crossover from 
 the superconducting state to the insulating state takes place. 
 In fact, in the experiment \cite{haviland98}, the behaviors of resistance 
 for finite $N$'s (63,127,255) change around $J_0\sim 0.5$, so we can state 
 that the transitions occur around there. 
 The positions of critical regions are slightly dependent on $N$. To be more 
 specific, with increasing $N$, the crossover is seen to occur at larger $J$. 
 This seems to correspond to the ascent of the transition temperature due to 
 the size effect in the case of a 2D $XY$ model, which can be detected 
 experimentally in superfluid $^4{\rm He}$ in mesoporous media 
 \cite{shirahama90,yano98}.

 In our theoretical derivation, when the zero point fluctuation of $\kappa$ 
 and it's potential barrier are comparable, the phase cannot be localized and 
 the phase coherence is destroyed. So we can define this configuration as the 
 boundary between superconducting and insulating states. The critical value 
 $J_{\rm C}$ is estimated from 
 Eq.\ (\ref{eq:potential}) and Eq.\ (\ref{eq:zero point frequency}), 
	\begin{equation}
	 J_{\rm C} \sim \left\{ (N-1)-(N-3)\cos\frac{\pi}{N-3} \right\}^{-1}.
	 \label{eq:SI transition point}
	\end{equation}
 For large $N$, Eq.\ (\ref{eq:SI transition point}) reduces to 
	\begin{equation}
	 J_{\rm C} \sim \frac{1}{2} \left\{ 1-\frac{1}{4} 
	 \left( \frac{\pi}{N}\right)^2 \right\}.
	 \label{eq:SI transition point,N infty}
	\end{equation}
 Eq.\ (\ref{eq:SI transition point,N infty}) qualitatively explains the size 
 dependence of the critical region in the experiment, although the quantitative  agreement is rather poor. 
 This quantitative discrepancy may result from the harmonic approximation we 
 have employed for estimating zero-point fluctuation of $\kappa$. 

 Now, we give some comments to our argument up to this point.
 There might be an assertion that we should take into account the translational  symmetry about the position of a phase slip center and therefore multiply 
 the tunneling rate by the number of virtual centers of a phase slip formation,  say, the number of junctions. But the mode we have adopted as the zero point 
 fluctuation of $\kappa$ is created by the collective motion of the field and 
 not localized at some junction, so the degree of freedom about the position of  a phase slip center is already summed up. 
 Besides, it is worth considering additional tunneling paths slightly shifted 
 from the valley paths which we have adopted. This supplement will be crucial 
 when the effective inertial mass of a phase slip is greatly dependent on the 
 tunneling path and the correction by these paths probably increase the 
 tunneling rate. 
 On the contrary, it is widely known that the coupling of superconducting 
 component with the environments generally suppresses the tunneling rate 
 \cite{caldeira81}, which we have ignored in this Letter. To evaluate the 
 contribution from this dissipation effect is also left for the future work.

 In conclusion, we have studied the quantum nucleation process of a phase slip 
 in a 1D JJ array using the collective coordinate method. 
 The theoretical value of the zero bias resistivity calculated by means of WKB 
 method in the case of large $N$ limit is in good 
 agreement with the experimental data asymptotically in the superconducting 
 side. We also have estimated the S-I transition point for an array 
 with a general number of junctions, which roughly agrees with the experiment.
 All of our results suggest that the intrinsic resistivity in the 
 superconducting side in a 1D JJ array at zero temperature can be understood 
 with the concept of quantum nucleation of phase slips and it's value can be 
 actually estimated.
 \\ \\

 This work is supported by Grant-in-Aid for Scientific Research from 
 the Ministry of Education, Science, Sports, and Culture of Japan.


\end{document}